\documentstyle[twocolumn,prb,psfig,aps,floats]{revtex}
\begin{document}
\sloppy \raggedbottom
 \draft \wideabs{
\title{Dimensional crossover and hidden incommensurability
in Josephson junction arrays of periodically repeated Sierpinski
gaskets}
\author{R. Meyer$^{1}$, S.E. Korshunov$^{2}$, Ch. Leemann$^{1}$, and P.
Martinoli$^{1}$}
\address{$^{1}$Institut de Physique, Universit\'{e} de Neuch\^{a}tel,
CH-2000 Neuch\^{a}tel, Switzerland}
\address{$^{2}$L.D. Landau Institute for Theoretical Physics,
Kosygina 2, 117940 Moscow, Russia}
\date{\today}

\maketitle

\begin{abstract}
We report a study of overdamped Josephson junction arrays with the geometry
of periodically repeated Sierpinski gaskets. These model superconductors
share essential geometrical features with truly random (percolative) systems.
When exposed to a perpendicular magnetic field $B$, their euclidian or
fractal behavior depends on the relation between the intervortex
distance (imposed by $B$) and the size of a constituent gasket, and
was explored with high-resolution measurements of the
sample magnetoinductance $L(B)$. In terms of the frustration
parameter $f$ expressing (in units of the superconducting flux
quantum) the magnetic flux threading an elementary triangular cell
of a gasket, the crossover between the two regimes occurs at
$f_{cN}=1/(2\times4^{N})$, where $N$ is the gasket order. In the
fractal regime ($f>f_{cN}$) a sequence of equally spaced
structures corresponding to the set of states with unit cells not
larger than a single gasket is observed at multiples of $f_{cN}$,
as predicted by theory. The fine structure of $L(f)$ radically
changes in the euclidian regime ($f<f_{cN}$), where it is determined by
the commensurability of the vortex lattice with the effective potential
created by the array. Anomalies observed in both the periodicity and the
symmetry of $L(f)$ are attributed to the effect of a hidden incommensurability,
which arises from the deformation of the magnetic field distribution
caused by the asymmetric diamagnetic response of the superconducting
islands forming the arrays.
\end{abstract}
}

\section{Introduction}

A wide variety of disordered materials, including superconductors,
is known to exhibit geometrical inhomogeneities over a broad range
of length scales. The properties of such systems can be
conveniently described in terms of percolation \cite{BH}, the
simplest idea to understand randomness. Percolation can be
regarded as a geometrical phase transition taking place at a
"critical concentration" $p_{c}$ separating a phase of finite
clusters ($p<p_{c}$) from a phase where an infinite cluster is
present ($p>p_{c}$). Like other critical phenomena, it is
characterized by a correlation length $\xi_{p}(p)$ which diverges
at the percolation threshold $p_{c}$. Right at $p_{c}$, a system
with percolative disorder exhibits a natural self-similar
structure at all length scales and can therefore be modeled by a
family of scale-invariant fractal lattices, such as the Sierpinski
gasket (SG), which has been suggested \cite{GAMK} to mimic the
{essential} geometrical features of the percolating cluster's
backbone. In the critical region above and below $p_{c}$, where
$\xi_{p}$ is finite, the nature of the geometry depends on the
length scale $\textit{l}$ at which one is probing the system: if
$\textit{l}<\xi_{p}$, its structure is \textit{fractal}, whereas
it can be regarded as homogeneous with conventional
\textit{euclidian} features for $\textit{l}>\xi_{p}$.

Allowing an accurate control of both the nature and the amount of
disorder and exhibiting properties quite sensitive to
dimensionality, Josephson junction arrays and wire networks
prepared with modern micro- and nanofabrication techniques provide
ideal model systems to investigate the dimensional crossover from
the euclidian to the fractal regime. The first step in this
direction was made by Gordon {\em et al.} \cite {GG1}, who
investigated the superconducting-to-normal phase boundary
$T_{c}(B)$ of aluminum wire networks formed by periodically
repeated SGs and of analogous networks with percolative geometry
exposed to a perpendicular magnetic field $B$. In those
experiments the magnetic length,
$\textit{l}(B)\approx(\phi_{0}/B)^{1/2}$, which is a measure of
the typical nearest-neighbor distance between the vortices present
in the system, was shown to be the relevant length scale to
explore the euclidian-fractal (EF) crossover ($\phi_{0}$ is the
superconducting flux quantum). While the scaling behavior of the
phase boundary of the SG networks was found to exhibit a crossover
from the euclidian to the fractal regime consistent with
theoretical predictions \cite{RT} based on extensions \cite{dG,A}
of the Ginzburg-Landau theory and allowing for comparison
\cite{GG3} with the anomalous diffusion exponent, no EF crossover
was observed in percolative networks which, surprisingly, were
found to behave like a homogeneous system at all length scales
$\textit{l}(B)$ probed in the experiment.

More recently, compelling evidence for the EF crossover in a percolative system
emerged from ac conductance measurements \cite {EAK} performed in
zero magnetic field on (unfrustrated) site-diluted Josephson
junction arrays with site occupation probabilities $p$ very close
to $p_{c}$. In these experiments the crossover was controlled by
the driving angular frequency $\omega$, which determines the ratio
of the impedances associated with the two types of links forming
the random network. According to Efros and Shklovskii \cite{ES}, the increase
of this ratio with decreasing $\omega$ also leads, if $p$
is sufficiently close to $p_{c}$, to a crossover from the fractal to
the euclidian regime, as confirmed by the experiments of Ref. \onlinecite {EAK}.

In this article we report the results of experimental and theoretical
studies of a model superconductor sharing essential geometrical features
with a percolative system. The samples we
have investigated are arrays of proximity-effect coupled SNS
junctions (where S stands for superconductor and N for normal
metal) consisting of SGs connected to each other at the vertices
in such a way as to form a regular triangular lattice. As can be
seen from Fig. 1(a), where a part of an array of second-order
gaskets is shown, in these systems the linear size $L_{N}=2^{N}a$
of an individual gasket of order $N$ can be regarded as playing
the role of $\xi_{p}$ ($a$ is the length of an elementary link of
the gasket). As shown in detail in this work, a remarkable feature of
the SG arrays is that in these systems the EF crossover is clearly manifest,
in contrast to truly percolative systems, where its signatures are elusive.

\begin{figure}[tb]
\vbox{ \centerline{
\psfig{file=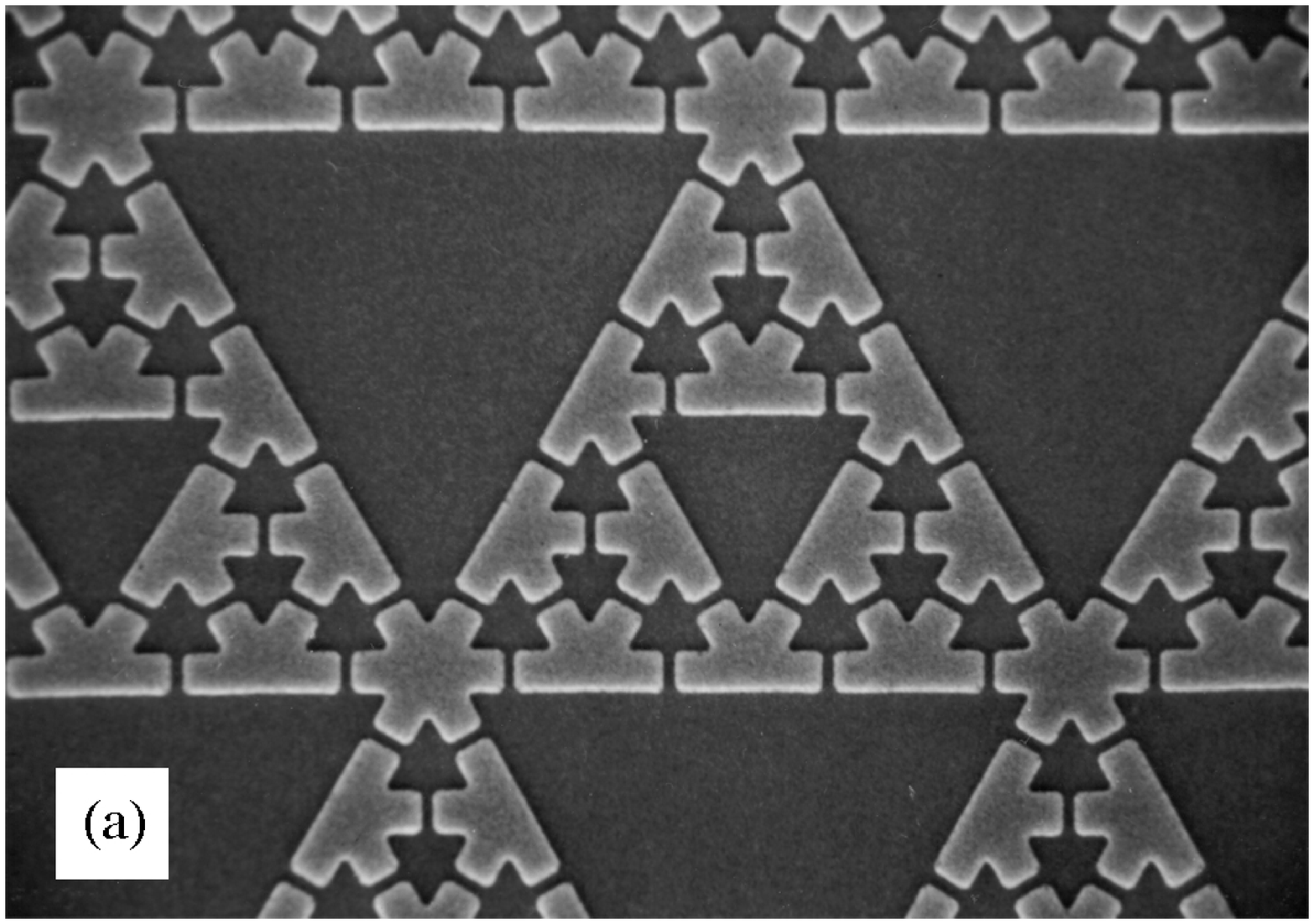,width=0.6\columnwidth,clip=}} \centerline{
\psfig{file=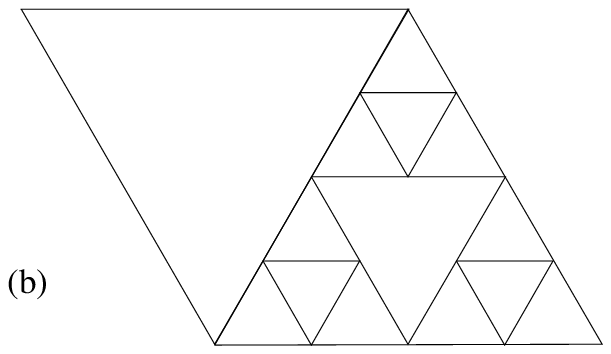,width=0.5\columnwidth,clip=}} \vspace{1mm}
\caption{(a) Scanning electron micrograph showing a portion of a
triangular array of periodically repeated second-order Sierpinski
gaskets of proximity-effect coupled Pb/Cu/Pb Josephson junctions.
The length of the elementary links of the gaskets is $8 \mu m$.
Notice the "truncated-star" shape of the superconducting Pb
islands (with the exception of those centered at the common
vertices of three constituent gaskets). (b) The rhombohedral unit
cell of a periodic array of second-order gaskets.} \label{fig1} }
\end{figure}

The quantity at the heart of the present study is the magnetoinductance
$L(B)$ of the SG arrays, extracted from measurements of their ac impedance.
Its interest resides in the observation that, being inversely proportional
to the areal superfluid density, it provides a tool to appreciate how
the degree of superconducting phase coherence in the system
changes with $B$ or, equivalently, with the level of frustration
imposed by $B$. Previous impedance measurements performed on
weakly frustrated arrays similar to those studied in this work
focused merely on the fractal regime and demonstrated, in
particular, the unusual scaling properties \cite{VKB} of the
vortex energy \cite{MGJ} as well as the asymptotic ($B\rightarrow
0$) scaling behavior of the field-induced correction to the array
inductance resulting from the hierarchical structure of the
gaskets \cite{KMM}.

Although some preliminary evidence for a
dimensional crossover was already reported in Ref.
\onlinecite{KMM}, the phenomenon was not exhaustively
investigated. In the present work we rely on high-resolution
studies of the complex fine structure of $L(B)$ to explore in
detail both the euclidian and the fractal regimes of the SG
arrays. Reflecting flux quantization phenomena occurring in loops
with a hierarchical distribution of sizes up to the gasket size
$L_{N}$, the fine structure provides a unique tool to reveal how
the geometrical properties of the system change as the magnetic
length $\textit{l}(B)$ is swept through $L_{N}$. We show that, in
terms of the frustration parameter $f$ expressing the magnetic
flux threading an elementary triangular cell of a gasket in units
of $\phi_{0}$, the EF crossover occurs at
$f_{cN}=1/(2\times4^{N})$. In the fractal regime  ($f>f_{cN}$) the
most relevant contributions to the fine structure of $L(f)$ are
shown to arise from a particular set of ground states defined by
$f=Mf_{cN}$, where $M$ is an integer. Corresponding to vortex
configurations where the vortex lattice is strongly pinned by
the hierarchical potential landscape created by the gaskets, these
states are particularly robust against thermal fluctuations and
are therefore quite prominent in the fine structure of $L(f)$.

A very interesting aspect emerged from the study of the
magnetoinductance in the fractal regime. The analysis of the data
revealed anomalous features (specifically, the suppression of
the periodicity corresponding to a shift of $f$ by $1$ and of the
symmetry with respect to $f=1/2$) inconsistent with theoretical
predictions based on the description of the system in terms of a
uniformly frustrated $XY$ model. We suggest that, because of the
asymmetric shape of the superconducting islands forming the
junction pattern of a gasket [see Fig. 1(a)], the screening
currents flowing in the islands create a distortion of the
magnetic field distribution in the array such that the fluxes
threading the various loops slightly deviate from being
proportional to their areas. This introduces an effective
incommensurability (which we call "hidden" to distinguish it from
the "geometric" one studied earlier \cite {BSN} in wire
networks with incommensurate cells) and perturbs the
self-similarity of the gaskets. As a result, the system is no
longer uniformly frustrated. We demonstrate that the
anomalous features mentioned above can be quantitatively accounted
for by a simple model, in which the area of the different
plaquettes of a gasket is changed according to an appropriate
deformation scheme.

Compelling evidence for the existence of the euclidian regime is
provided by the array's magnetoinductance for $f<f_{cN}$. Besides
the absence of the power-law scaling behavior characteristic of
the fractal regime, $L(f)$ contains structures which reflect the
presence of ground states corresponding to vortex configurations
commensurate with the underlying triangular lattice formed by the
largest triangular loops of the array (below $f_{cN}$, it is
energetically unfavorable for vortices to penetrate loops of
smaller size), thereby allowing an unambiguous identification of
the euclidian regime.

In general, these vortex configurations have the same structure as in the $XY$
model on a honeycomb lattice, but correspond to values of $f$
reduced by a factor of $1/f_{cN}=2\times 4^N$. However, the case
$f=(1/2)f_{cN}$ requires special attention. In a honeycomb lattice
the ground state of the corresponding (fully frustrated) $XY$
model is characterized by an accidental degeneracy, which (to lowest
order) survives even in the presence of thermal fluctuations \cite{K}. Owing
to the more complex structure of the system, however, this peculiar degeneracy
is removed in our SG arrays, allowing to identify the vortex
configuration in the ordered phase at $f=(1/2)f_{cN}$.

The paper is organized as follows.
Experimental details are given in Sec. II. Sec. III is devoted to
the fractal regime ($f>f_{cN}$). Relying on the methods developed
in Ref.\onlinecite{KMM}, in Sec. III A we present the calculation
of the magnetoinductance of a frustrated SG array for the
particular set of frustrations $f=Mf_{cN}$ corresponding to the
sequence of the most stable states, which are characterized by a
relatively compact structure (with a unit cell not larger
than a single gasket). Experimental data for the magnetoinductance
in the same regime are presented and discussed in Sec. III B,
where we show, in particular, how the anomalous features of
$L(f)$, revealing the presence of hidden incommensurability, can
be accounted for by a simple model, in which the areal changes of
different plaquettes (related to the redistribution of the
magnetic field) are determined by only one adjustable parameter.
In Sec. IV we provide experimental evidence for the existence of
the EF crossover and in Sec.V the fine structure of $L(f)$ in the
euclidian regime below $f_{cN}$ is shown to be consistent with the
existence of vortex states commensurate with the periodic lattice
formed by the largest triangular cells of the arrays. A few
concluding remarks are given in Sec. VI.

\section {Experimental aspects}

The samples studied in this work consist of second- ($N=2$) and
fourth-order ($N=4$) gaskets sitting on the sites of,
respectively, a $313\times313$ and a $78\times78$ triangular
lattice and connected to each other at the vertices [see Fig.
1(a)]. Each gasket contains, respectively, $3^{2+1}=27$ and
$3^{4+1}=243$ SNS Josephson junctions consisting of
superconducting Pb islands proximity-effect coupled to each other
by an underlying normal Cu layer. The geometrical and physical
parameters of the junctions are almost identical to those of the
array studied in Ref. \onlinecite{KMM}. Most of the data presented
below have been obtained in experiments performed on the array of
second-order gaskets.

The sheet magnetoinductance $L(f)$ was
inferred from measurements of the array's {\em linear} sheet
impedance $Z=R+i\omega L$ performed with a very sensitive
SQUID-operated two-coil mutual inductance technique \cite{JGM} at
driving frequencies typically in the range 0.1-1.0 kHz. With this
method we were able to resolve inductance changes of the
order of 10 pH in swept-frustration measurements, in which $f$
could be tuned with a precision better than $10^{-3}$. The
experimental data are presented and analyzed in terms of
$L^{-1}(f)$, the quantity measuring the degree of superconducting
phase coherence in the samples. When needed, the resistive
component $R(f)$, related to dissipative vortex motion, is also
shown for completeness. Additional details concerning the samples
and the measuring technique can be found in Ref. \onlinecite{KMM}.

In the following, temperatures are expressed in terms of the
reduced temperature relevant for the statistical mechanics of the
system, $\tau\equiv k_{B}T/J(T)$, where $J(T)$ is the
temperature-dependent Josephson coupling energy.  At temperatures
well below the zero-field critical temperature ${\tau_{cN}}$,
$J(T)$ was deduced from measurements of the "bare" sheet
kinetic inductance
${L(T)=(\phi_{0}/2\pi)^{2}(5/3)^{N}/\sqrt{3}J(T)}$ of the
unfrustrated samples \cite{KMM}.  Extrapolation to higher
temperatures was then achieved by fitting the low-temperature
data to theoretical expressions \cite{ZZ} for ${J(T)}$.

Because of their two-dimensional (2D) nature at length scales larger than
$L_{N}$, both samples are expected to exhibit, at zero frustration
($f=0$) and in the limit $\omega \rightarrow 0$, a
Berezinskii-Kosterlitz-Thouless (BKT) phase transition \cite{BKT}.
A sharp depression of $L^{-1}(0)$, which can be associated
with the BKT transition, has been indeed observed at,
respectively, $\tau_{c2}\approx 0.57$ and $\tau_{c4}\approx0.23$,
in good agreement with the theoretical prediction \cite{VKB}
$\tau_{cN}=(3/5)^{N}\tau_{c0}$, where $\tau_{c0}\approx1.5$ is the
reduced temperature of the BKT transition of a regular triangular
Josephson junction array \cite{SS} with the same $J(T)$.

\section{The fractal regime}

\subsection{The ground states of a regular array of Sierpinski gaskets
and their sheet inductance}

We start by recalling that, within the
framework of an approximation ignoring thermal fluctuations,
a Josephson junction array behaves, with respect to an
external (dc) current source, like a network of inductors
\{$L_{ij}$\}, whose inductances are given \cite{KMM,YS} by:
\begin{equation}
L_{ij}(\theta_{ij})=\frac{(\phi_{0}/2\pi)^2}{J\cos\theta_{ij}}\hspace{5mm},
\label{eq1}
\end{equation}
where $\theta_{ij}$ is the gauge-invariant phase difference across
the link $ij$. As required by fluxoid quantization, the sum of
$\theta_{ij}$ around a lattice cell is equal to $2\pi (fS-m)$, where
$f$ is the frustration parameter expressing the magnetic flux (in units
of $\phi_{0}$) threading an elementary triangular cell of a gasket
[$f=Ba^{2}\sqrt{3}/(4\phi_{0})$], $S$ the area of the cell (expressed in units of
the area of an elementary triangular loop) and $m$ the number of flux quanta
(vortices) penetrating the cell under consideration.

In writing Eq.  (1) we have assumed that the proximity-effect coupled SNS
junctions forming the arrays studied in this work have a sinusoidal
current-phase relation at the temperatures of interest \cite {L}. It clearly
follows from Eq. (1) that, even if all the junctions are identical, their
effective inductances in a frustrated system may differ substantially from
each other on account of the nonuniform distribution of
$\{\theta_{ij}\}$. Since the array magnetoinductance $L(f)$ can be
found by applying Kirchhoff's laws to the inductor network
\{$L_{ij}$\}, it is evident that, at any frustration $f$, $L(f)$
will be completely determined once the distribution of
$\{\theta_{ij}\}$ is known.

The ground state of a uniformly frustrated array of periodically
repeated gaskets of order $N$ [see Fig. 1(a)] can be constructed
by a simple juxtaposition after finding the ground state of an
isolated $N$th-order gasket only if the constraints of fluxoid
quantization imposed on the largest triangular loops (located
between the $N$th-order gaskets) are automatically fulfilled.
Recalling the definition of $f$ given above, it can be shown
\cite{KMM} that this condition is satisfied only for a particular
set of frustrations given by:
\begin{equation}
f=\frac{M}{2\times4^{N}}\hspace{5mm}, \label{eq2}
\end{equation}
where $M$ is an integer corresponding to the total number of
vortices in the rhombohedral unit cell of the SG array composed,
as shown in Fig. 1(b), by a single $N$th-order gasket and the
adjacent "empty" triangular loop. Thus, in order to calculate the
magnetoinductance of the system at the values of $f$ given by Eq.
(2), all we need is to determine the ground-state distribution of
$\{\theta_{ij}\}$ in one of its constituent gaskets.

\begin{figure}[tb]
\vbox{ \centerline{ \psfig{file=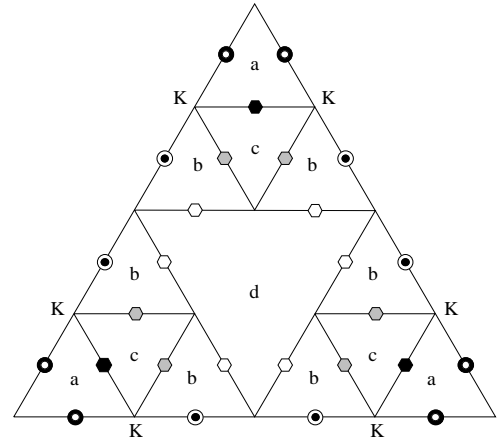,width=0.8\columnwidth
,clip=}} \vspace{2mm} \caption{The links of this second-order
gasket sharing the same symbol have identical gauge-invariant
phase differences $\theta_{ij}$ in the states whose symmetry is
consistent with the reflection and third-order rotation symmetries
of the gasket. The 5 independent $\{\theta_{ij}\}$ follow by
imposing current conservation at one of the 6 equivalent nodes $K$
and fluxoid quantization in the 4 non-equivalent loops $a$, $b$,
$c$ and $d$.} \label{fig2} }
\end{figure}

Assuming that the ground state of a gasket is the one having the highest
possible symmetry consistent with its reflection and third-order
rotation symmetries \cite {AH} (for wire networks with the
geometry of a third-order SG this conjecture has been confirmed by
numerical calculations \cite{CD}), it can be shown \cite {M} that
the number of independent bond variables $\{\theta_{ij}\}$ is
equal to $(3^{N}+1)/2$. They can be found from an equal number of
constraints imposed by current conservation at the nodes and
fluxoid quantization in the loops of the gasket.  For example, in
the second-order gasket of Fig. 2 the links sharing the same
symbol have the same values of $\theta_{ij}$, thereby showing that
there are only 5 independent bond variables.

Imposing current conservation at one of the 6 equivalent nodes (denoted
by $K$ in Fig. 2) and fluxoid quantization in the 4 non-equivalent loops
$a$, $b$, $c$, and $d$ (with, respectively, vortex occupation
numbers $m_{a}$, $m_{b}$, $m_{c}$, and $m_{d}$) leads to a system
of 5 equations (only one of which, describing current
conservation, is nonlinear) which have to be solved numerically
under the constraint that the distribution of the quantum numbers
$\{m_{\alpha}\}$ ($\alpha=a,b,c,d$) is such that the gasket energy:
\begin{equation}
E=J\sum_{<ij>}(1-\cos\theta_{ij})\; \label{E}
\end{equation}
is minimized. Quite remarkably, for a given frustration satisfying Eq.
(2), the distribution of $\{m_{\alpha}\}$ corresponding to the lowest
energy turns out to be identical \cite{M} to that emerging from a
calculation based on junctions with a linear current-phase
relation, for which a fully analytical treatment is possible \cite
{AH,CD}. The result is illustrated in Fig. 3, where the
ground-state vortex configurations for the rhombohedral unit cell
of a regular array of second-order gaskets are shown for $M$ in
the range [$0,40$].

Inspection of Fig. 3 reveals characteristic
features, which are valid for arbitrary gasket order.  One first
observes that, with increasing frustration, vortex nucleation
spreads from the largest to smallest loops \cite{MGJ,CD}, a
property reflecting the hierarchical character of the energy
needed to create a vortex excitation \cite{VKB,MGJ}. Next, one
notices that the vortices penetrate the gaskets only for $M>1$,
thereby implying that a SG array will exhibit fractal behavior
only for $f>1/(2\times4^{N})$. One further recognizes that, since
for $M=1$ the rhombohedral unit cell contains just one single
vortex sitting in the largest triangular loop, the ground state of
the array at $f=1/(2\times4^{N})$ corresponds to a triangular
lattice of vortices with a nearest-neighbor distance equal to the
gasket size $L_{N}$. Recalling that $L_{N}$ plays the role of
$\xi_{p}$, one expects that for $f<1/(2\times4^{N})$ the system
will behave like a regular Josephson junction array with
conventional euclidian geometry. Thus, we identify
\begin{equation}
f_{cN}=\frac{1}{2\times4^{N}}\hspace{5mm} \label{eq3}
\end{equation}
as the frustration at which the EF crossover occurs. The nature
of the ground states of the SG array in the euclidian regime
($0<f<f_{cN}$) is discussed in Sec.  V.

\begin{figure}[tb]
\vbox{ \centerline{
\psfig{file=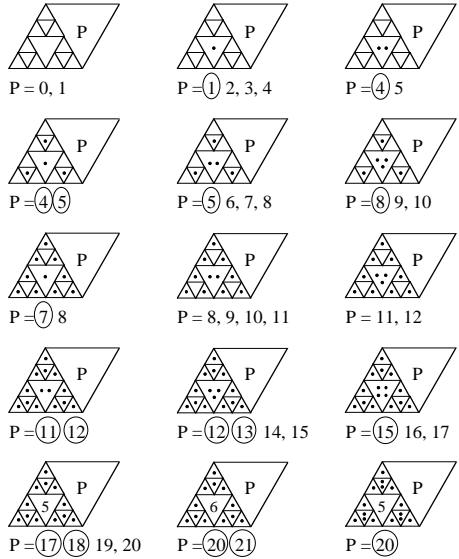,width=0.7\columnwidth,clip=}} \vspace{3mm}
\caption{Ground-state vortex configurations at multiples of
$f=1/32$ in the range [$0,40$] for the rhombohedral unit cell of a
uniformly frustrated periodic array of (undistorted) second-order
gaskets. $P$ denotes the number of vortices in the largest
triangular loop. Vortex configurations with circled vortex
occupation number $P$ are no longer ground states of the weakly
distorted gasket of Fig. 6(a) (see Fig. 7 for a comparison).}
\label{fig3} }
\end{figure}

Having shown how the
structure of the ground state can be determined for the particular
set of frustrations (2), we can now proceed with the calculation
of the \textit{sheet} magnetoinductance $L(f)$ of the SG array,
the quantity measured in our experiments. We first notice that,
for this particular set of frustrations, $L(f)$ is proportional to
the magnetoinductance of a constituent gasket. Therefore, if we
calculate the inductance of a single gasket and normalize it to
its value at $f=0$ to eliminate the trivial dependence on the
gasket size, we obtain a result also expressing the normalized
sheet magnetoinductance $L(f)/L(0)$ of the composite periodic
system. In Ref. \onlinecite{KMM} we have pointed out that for a
given distribution of $\{\theta_{ij}\}$ the calculation of the
inductance of a single gasket
can be performed by successive application of the triangle-star
transformation well known in the theory of electric networks
\cite{SW}. The result of such a calculation for a regular array of
second-order gaskets is shown in Fig. 4 for multiples of $f=1/32$ in the
range $[-11,43]$. In order to compare this calculation with the
experimental data presented in the following subsection, we plot
the normalized inverse magnetoinductance $L^{-1}(f)/L^{-1}(0)$. Notice
that, as expected for a uniformly frustrated Josephson junction array,
$L^{-1}(f)$ is symmetric with respect to $f=1/2$ and periodic with period $f=1$.
We also recall that, although hardly visible in the linear plot of
Fig. 4, in the fractal regime the frustration-induced correction
$\Delta L(f)=L(f)-L(0)$ to the array inductance is predicted to
scale \cite{KMM}, in the limit of small frustrations and low
temperatures, as $f^{\nu}$ with $\nu=\ln(125/33)/\ln4\approx0.96$.
Obviously, the power-law behavior of $\Delta L(f)$ should no
longer persist in the euclidian regime below $f_{cN}$.

\begin{figure}[tb]
\vbox{ \centerline{
\psfig{file=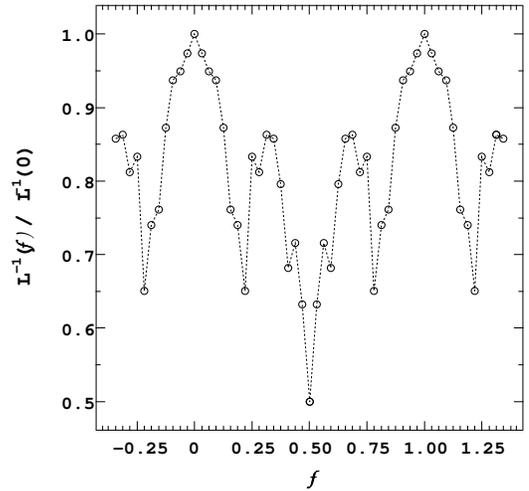,width=0.8\columnwidth,clip=}} \vspace{1mm}
\caption{Normalized inverse magnetoinductance at multiples of
$f=1/32$ in the range [$-11,43$] of a uniformly frustrated regular
array of (undistorted) second-order gaskets. Notice the symmetry
with respect to $f=1/2$ and the periodicity of period $f=1$. The
dotted line is simply a guide to the eye.} \label{fig4} }
\end{figure}

So far we have considered only the frustrations given by Eq. (2), for which
the ground state of a regular array of gaskets can be regarded as
a periodic replication of the ground state of a single gasket. In
order to determine the array ground state at  rational
frustrations differing from those given by Eq. (2), one should
consider a "supercell" comprising more than one gasket. The
analysis of the ground states based on such supercells rapidly
becomes cumbersome and (for $f>f_{cN}$) is not pursued in this
work. However, as supercells imply that superconducting phase coherence
extends at larger length scales, one can expect the
corresponding ground states to be more vulnerable to thermal
fluctuations [and, consequently, less prominent in the fine
structure of $L(f)$] than those at $f=Mf_{cN}$. At these
particular values of frustration the vortex configurations (shown
in Fig. 3) are strongly pinned by the hierarchical potential
landscape provided by the gaskets \cite{MGJ}, thereby making these
ground states particularly robust against thermal fluctuations.

\subsection{Comparison with experiment and effects of hidden
incommensurability}

Focusing on the fine structure of $L(f)$ we now compare the
theoretical predictions of Sec. III A with high-resolution
magnetoinductance measurements performed on the array of
second-order gaskets. Fig. 5(a) shows the normalized inverse sheet
magnetoinductance $L^{-1}(f)/L^{-1}(0)$ of the array of
second-order gaskets (measured at 1 kHz) at three different
(reduced) temperatures. We first observe that, although the
overall shape of the magnetoinductance curves looks roughly
similar, the fine structure becomes richer and much sharper with
increasing temperature, thereby revealing very clearly almost all
the "superfluid" peaks corresponding to the states with a unit
cell not larger than a single gasket and belonging to the sequence
$f=M/32$ given by Eq. (\ref{eq2}). Notice that, to make the
identification of the structures easier, the frustration unit on
the horizontal axis of Fig. 5(a) is chosen to be equal to
$f_{cN}=1/32$. The striking evolution of the fine structure with
temperature suggests that the motion of vortices due to thermal
fluctuations plays a major role in the dynamic response of these
arrays.  We interpret the behavior shown in Fig. 5(a) as clear
evidence that, at sufficiently high temperatures, superconducting
phase coherence in the neighborhood of the ground states at
$f=M/32$, for which the vortex lattice is pinned, is drastically
disrupted by vortex-lattice defects, created by excess or missing
vortices, moving almost freely on the "frozen" vortex background.
This process dramatically sharpens the fine structure, thereby
enhancing the amplitude of the oscillations. Similar behavior was
also observed in experiments performed on wire networks of
interconnected gaskets \cite{MGJ} and in regular triangular
Josephson junction arrays \cite{TSM} as well as in numerical
simulations \cite{YS}.

In sharp contrast to the theoretical prediction for a uniformly frustrated
SG array (see Fig. 4), the $L^{-1}(f)/L^{-1}(0)$ curves of Fig. 5(a), although
still symmetric with respect to $f=0$, are no longer periodic with
period $1$, a behavior leading unavoidably to the suppression of
the symmetry with respect to $f=1/2$. We attribute these anomalous
features to the \textit{inhomogeneous frustration} resulting from
the change in the effective areas of different plaquettes caused
by the asymmetric (with respect to the direction of the links)
diamagnetic response of the truncated-star-shaped superconducting
(Pb) islands [see Fig. 1(a)]. Because of this particular
geometrical form, the screening currents flowing in these grains
create a distortion of the current patterns associated with the
individual loops which leads to a redistribution of the magnetic
field and perturbs the self-similarity of the gaskets. In the
temperature range ($5.5 K<T<6.4 K$) of the data shown in Fig. 5(a), the
magnetic penetration depth of the Pb islands [as estimated from the
zero-temperature bulk Pb value ($\lambda(0)\approx 40 nm$) and the
(proximity-effect reduced) transition temperature ($T_{c}=6.9 K$) of the
Pb grains] is at least 25 times smaller than their smallest planar geometrical
dimension, which corresponds to the width of the junctions
($\simeq 2 \mu m$). Thus, one expects the distortion of the
current patterns and, consequently, the nonuniformity of the
frustration to have a considerable effect on $L^{-1}(f)$.

\begin{figure}[tb]
\vbox{ \centerline{
\psfig{file=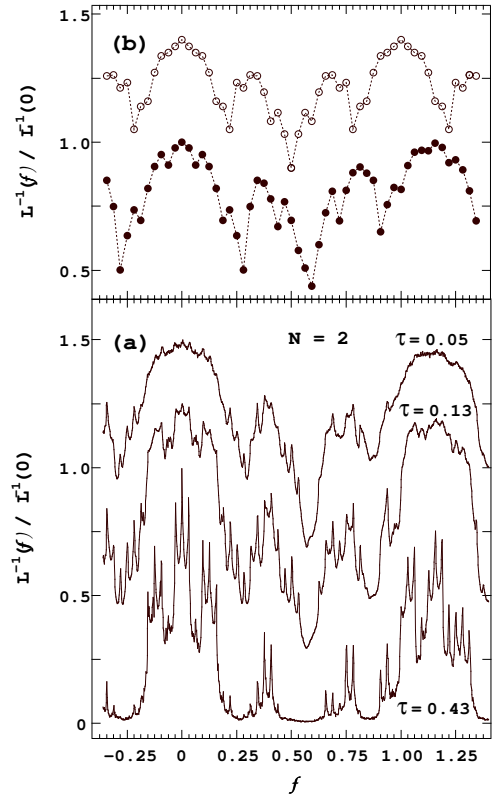,width=0.75\columnwidth,clip=}} \vspace{2mm}
\caption{(a) Normalized inverse magnetoinductance of the periodic
array of second-order gaskets shown in Fig. 1(a) measured at 1 kHz
at three different temperatures as a function of frustration. For
clarity, the curves for $\tau=0.13$ and $\tau=0.05$ are shifted
upwards by, respectively, 0.25 and 0.50. Notice the absence of
periodicity of period $1$ and of the related symmetry with respect
to $f=1/2$. (b) Solid circles: normalized inverse
magnetoinductance calculated for an array of second-order gaskets
deformed as shown in Fig. 6 with $\epsilon=7.4\%$. This curve
should be compared with the data at the lowest temperature
($\tau=0.05$) in (a). For comparison, the calculation of Fig. 4
for an undistorted SG array is also shown (open circles, the curve
is shifted upwards by 0.4 for clarity). The dotted lines in (b)
are simply guides to the eye.} \label{fig5} }
\end{figure}

The origin of the phenomenon being intimately related to the
geometry of the superconducting islands rather than to the
physical properties of the junctions, it seems plausible to
describe the effect of the inhomogeneous frustration by changing
merely the effective area of the different plaquettes according to
a prescribed rule. This is illustrated in Fig. 6(a) for a
second-order gasket, whose distortion is modeled by shifting the
vertices of the triangular loops toward the "centers of mass" of
the corresponding superconducting islands, as shown in Fig. 6(b).
Since the distribution of the screening currents in the islands is
shifted in the same direction, this deformation scheme appears to
be a reasonable approach offering, above all, the advantage of a
simple description in terms of a single paramater, the (small)
areal change $\epsilon$ (expressed in units of the area of an
elementary triangular cell) defined in Fig. 6(a). Notice,
incidentally, that this procedure does not alter the frustrations
[Eq. (2)] for which the ground state of the array can be
constructed by a simple juxtaposition of independent gaskets.

\begin{figure}[tb]
\vbox{ \centerline{ \psfig{file=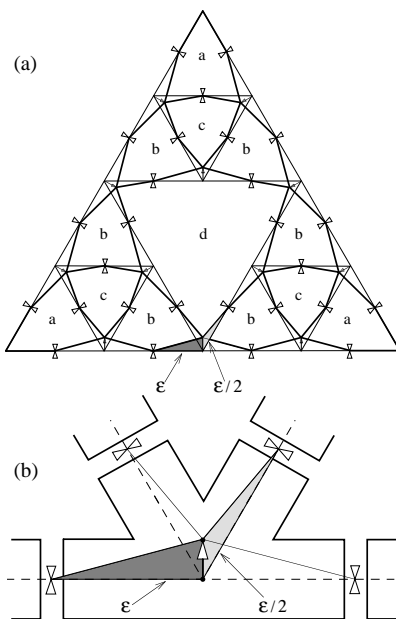,width=0.6\columnwidth
,clip=}} \vspace{5mm} \caption{(a) The deformation scheme (here
illustrated for a second-order gasket) introduced to describe the
effect of the nonuniform frustration resulting from the distortion
of the screening current pattern created by the asymmetric shape
of the superconducting islands. The vertices of the triangular
loops are shifted towards the "centers of mass" of the
corresponding superconducting grains as shown in (b). The areal
changes of the loops $a$, $b$, $c$ and $d$ are expressed in terms
of the deformation parameter $\epsilon$ (measured in units of the
area of an elementary triangular cell) defined in (b).}
\label{fig6} }
\end{figure}

As the resulting deformation preserves the reflection and third-order
rotation symmetries of the gasket [see Fig. 6(a)], the
determination of the vortex and $\{\theta_{ij}\}$ configurations
in the highly symmetric ground states at $f=Mf_{cN}$ can be
carried out by following again the procedure described in Sec. III
A. The only modification appears in the formulation of the
fluxoid-quantization constraints for the deformed loops, whose
areas (and, therefore, the associated magnetic fluxes) change. For
example, for the second-order gasket of Fig. 6(a) the areas of the
deformed plaquettes $a$, $b$, $c$, and $d$ are equal, in an approximation
linear in $\epsilon$, to $1-\epsilon$, $1-(3/2)\epsilon$,
$1-3\epsilon$, and  $4+3\epsilon$, respectively. An analogous problem,
however with a quite different deformation scheme, was considered
by Ceccatto {\em et al.} \cite{CD} for a system with a linear current-phase
relation.

\begin{figure}[tb]
\vbox{ \centerline{
\psfig{file=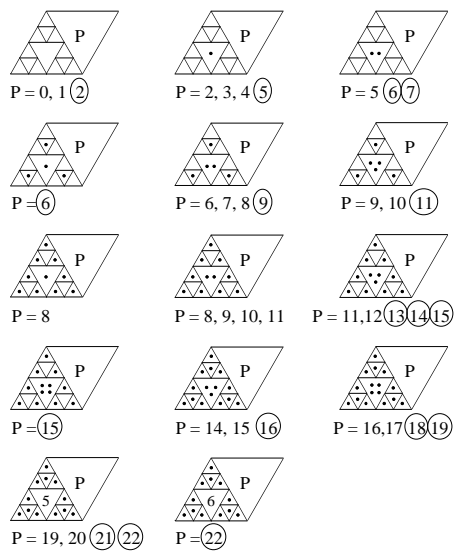,width=0.7\columnwidth,clip=}}
\vspace{3mm} \caption{Ground-state vortex configurations at
multiples of $f=1/32$ in the range [$0,40$] for the rhombohedral
unit cell of a periodic array of second-order gaskets deformed as
shown in Fig. 6(a) with $\epsilon=7.4\%$. $P$ denotes the number
of vortices in the largest triangular loop. The comparison with
the ground states of an array of undistorted gaskets of the same
order (Fig. 3) reveals that the vortex configurations with circled
vortex occupation number $P$ are new ground states of the system.}
\label{fig7} }
\end{figure}

Ground-state vortex configurations for the rhombohedral
unit cell of an array of second-order gaskets with a deformation
parameter $\epsilon=7.4 \%$ (the reason for the choice of this
value is explained below) are shown in Fig. 7 for $M$ in the range
[$0,40$]. A comparison with Fig. 3 reveals that some of the ground
states of the undistorted array have been replaced by new ones.
More precisely, while for the regular SG lattice the symmetry with
respect to $f=1/2$ implies that the ground state for $(1-f)$ may
be conceived as resulting from the superposition of the ground
states for $f=1$ and $-f$, for the distorted system this property
no longer holds. The absence of periodicity (corresponding to the
$f\leftrightarrow 1+f$ symmetry) and of the related
$f\leftrightarrow 1-f$ symmetry are clearly reflected in the
normalized inverse magnetoinductance shown in Fig. 5(b). This
$L^{-1}(f)/L^{-1}(0)$ curve was calculated for an array of
second-order gaskets using the method outlined in Sec. III A and
was fitted to the low-temperature ($\tau=0.05$) data of Fig.
5(a) using $\epsilon$ as an adjustable parameter. The best fit was obtained
for $\epsilon=7.4\%$. In this connection, it should be noticed that, although
weak, thermal fluctuations still affect the data at $\tau=0.05$ and tend
to enhance the amplitude of the oscillations with respect to that predicted
by our calculation, which neglects thermal fluctuations. Nevertheless,
the agreement is quite remarkable, especially if one considers that it involves
only one fitting parameter. Moreover, the vertex displacement
[denoted by the arrow in Fig. 6(b)] defining $\epsilon$ turns out
to be $\sim 70\%$ of the distance to the "center of mass" of the
truncated-star-shaped superconducting islands, thereby
demonstrating the basic validity of our interpretation.

Similar behavior was observed in the magnetoinductance of the array
of fourth-order gaskets, whose fine structure was found to be
richer than that of the array of second-order gaskets, as
demonstrated by the incipient splitting of some of the structures
at $f=M/32$. However, because of the strong overlap resulting
from the nonvanishing width of the superfluid peaks, only a
fraction of the states at multiples of $f=1/512$ could be resolved
and unambiguously identified in a plot at (relatively) low
frustration resolution like that of Fig. 5 (for high-resolution
data at very small $f$, see Fig. 9 in Sec. V).

Before closing this section, we would like to notice that when $\epsilon$ is a
rational number, the periodicity of $L^{-1}(f)$ is restored,
however with a period larger than 1.

\section {The crossover between the two regimes}

Below $f_{cN}$ [see Eq. (\ref{eq3})] the SG arrays are expected to
be in the euclidian (or homogeneous) regime. In order to provide
preliminary evidence for the existence of the EF crossover at
$f_{cN}$, in Fig.  8 we compare, in a log-log plot, the quantity
$\Delta L^{-1}(f)/L^{-1}(0)\equiv
[L^{-1}(0)-L^{-1}(f)]/L^{-1}(0)$, which measures the relative
change in superfluid density caused by frustration, for the two SG
arrays studied in this work. Both curves were taken at 160 Hz
and at temperatures such that the structures corresponding to the
ordered states are emphasized by thermal fluctuations. With
decreasing frustration the data for the sample of fourth-order
gaskets exhibit, down to $f_{c4}=1/512$, clear fractal features:
specifically, four self-similar stages (the number of stages being
consistent with the order of the gaskets) reflecting the
dilational symmetry of the gaskets and an overall scaling with $f$
which, in spite of the indisputable evidence for fluctuation
effects, follows the asymptotic prediction $\Delta L^{-1}(f)
\propto f^{\nu}$. For comparison, the result of a calculation for
an undistorted infinite gasket based on the methods discussed in
Sec. III A is also shown in Fig. 8. Below $f_{c4}$ the data tend
to flatten out, thereby signaling a possible change of regime.
However, considering the fact that this change sets in almost at
the limit of our inductance resolution ($\Delta L/L\sim 1\%$), we
refrain from drawing a firm conclusion with regard to the
existence of a dimensional crossover in the array of fourth-order
gaskets.

\begin{figure}[tb]
\vbox{ \centerline{
\psfig{file=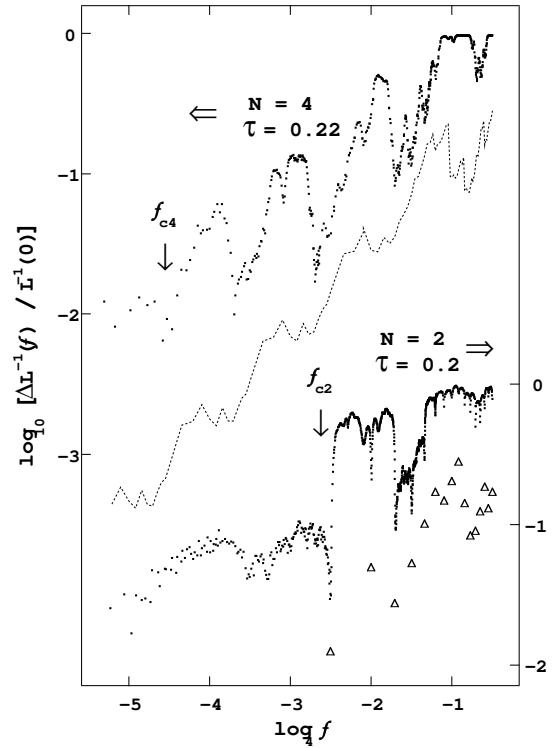,width=0.85\columnwidth,clip=}} \vspace{2mm}
\caption{Log-log plot of the frustration-induced relative change
of the inverse sheet inductance of the arrays of second-order
($N=2$) and fourth-order ($N=4$) gaskets measured at 160 Hz. The
dotted curve is the theoretical prediction for an undistorted
gasket of infinite order (see Ref. 12). The result [the same as in
Fig. 5(b)] of the calculation for an array of distorted
second-order gaskets is shown (open triangles) at multiples of
$f=1/32$ in the range [$1,16$]. For clarity, the theoretical
curves are shifted downwards by a quarter of a decade. $f_{cN}$ is
the frustration at which the crossover from the euclidian
($f<f_{cN}$) to the fractal ($f>f_{cN}$) regime occurs.}
\label{fig8} }
\end{figure}

On this subject, the data for the array of second-order
gaskets convey a much stronger message. In Sec. III B the
magnetoinductance of this sample was shown to obey the
predictions for the fractal regime whose signatures, although less
pronounced than in the $N=4$ case, can also be identified, above
$f_{c2}=1/32$, in the log-log plot of Fig. 8 [notice that the
self-similar and scaling properties of $\Delta L^{-1}$ are
expected to become clearly manifest only in the asymptotic limit
($f\rightarrow 0$) of higher-order gaskets \cite{KMM}]. Below
$f_{c2}$, however, the behavior of $\Delta L^{-1}(f)/L^{-1}(0)$
drastically changes. Besides the loss of self-similarity and
scaling, the data reveal, by closer inspection, the presence of
structures [the "dips" in $\Delta L^{-1}(f)$] corresponding to new
commensurate states, which can not be ascribed to the fractal
regime.

\begin{figure}[tb]
\vbox{ \centerline{
\psfig{file=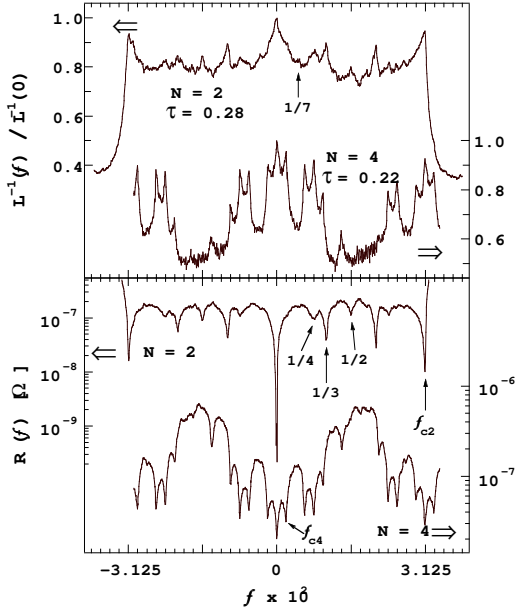,width=0.8\columnwidth,clip=}} \vspace{1mm}
\caption{Normalized inverse magnetoinductance and
magnetoresistance of the periodic arrays of second-order ($N=2$)
and fourth-order ($N=4$) gaskets measured at 10 Hz in the
frustration range $|f|\leq f_{c2}=1/32=0.03125$. Structures in the
data for $N=2$ are labeled in terms of $f_{H}=f/f_{c2}$, the
frustration parameter referred to the hexagonal unit cell of the
honeycomb lattice shown in Fig. 10. Structures at multiples of
$f_{c4}=1/512$ demonstrate the fractal character of the response
for $N=4$, whereas the prominent structures at $|f_{H}|=1/3$ and
the symmetry with respect to $f_{H}=1/2$ of the data for $N=2$ are
signatures of the euclidian regime.} \label{fig9} }
\end{figure}

To strengthen the evidence for a regime crossover, in Fig.
9 we present the results of sheet inductance measurements
performed at high frustration resolution in the range $|f|\leq
f_{c2}$. Once again, to promote structures corresponding to
ordered states we tuned the effect of thermal fluctuations by
increasing the temperature and reducing the measuring frequency
(10 Hz) in an appropriate way. Moreover, in Fig. 9 we also include
the dissipative component $R(f)$, whose remarkably well resolved
fine structure provides additional evidence for the EF crossover.
Notice that maxima in the (normalized) superfluid density,
$L^{-1}(f)/L^{-1}(0)$, correspond to minima in $R(f)$, as it
should be. For the array of fourth-order gaskets both the
superfluid and the dissipative components display clear fractal
features with marked structures at multiples of $f_{c4}=1/512$
corresponding to the frustration unit on the $f$-axis.
In sharp contrast with this behavior, the dynamic response of the
array of second-order gaskets is, within experimental accuracy,
symmetric with respect to $f=(1/2)f_{c2}$ and contains well
resolved structures at $|f|=(1/4)f_{c2}$, $|f|=(1/3)f_{c2}$ and
$|f|=(1/2)f_{c2}$. In particular, the presence of prominent
structures at $|f|=(1/3)f_{c2}$ provides compelling evidence for
the existence of a new regime, since no commensurate state
corresponding to this frustration can ever emerge from the
characteristic sequence $f=Mf_{cN}$ of the fractal regime for
gaskets of any order. We interpret these features as an
unambiguous signature of the euclidian regime, in which the fine
structure of the array's impedance reflects, as shown in the
following section, the existence of vortex ground states with a
unit cell larger than a second-order gasket.

\section {The euclidian regime}

In order to understand how our arrays of gaskets behave in
the euclidian regime, we first recall that the energy of a vortex
in a Sierpinski gasket decreases with increasing size
of the cell in which the vortex core is localized \cite{VKB,MGJ}.
Accordingly, a single vortex in a lattice of periodically repeated
SGs can be considered as interacting with an external potential
whose minima coincide with the centers of the largest triangular
cells located between the $N$th-order gaskets and, therefore, form
a triangular lattice. For $f<f_{cN}$ the number of vortices in the
system is smaller than the number of the largest triangular cells,
so that it is energetically favorable for the vortices to
occupy only these largest cells, the concentration of the occupied
ones being equal to the ratio $f/f_{cN}$. Therefore, one can
expect that, for $f<f_{cN}$, the behavior of an array of
periodically repeated SGs resembles that of a uniformly frustrated
$XY$-model on a \textit{honeycomb} lattice (with a reduced
value of frustration $f_H=f/f_{cN}<1$), whose ground states can
also be thought of as formed by vortices occupying the sites of a
triangular lattice with the same concentration $f_H$.

Another approach leading to the same conclusion relies on the
iterative procedure, based on successive applications of the triangle-star
transformation \cite{SW}, developed in Ref. \onlinecite{KMM}. At low
frustrations, such that there are no vortices in the gaskets, this method
allows to replace each gasket by an elementary "star" consisting of three
links with a modified interaction. This transforms a lattice of
periodically repeated SGs into a honeycomb lattice (see Fig. 10).
The difference with respect to the conventional $XY$-model on a
honeycomb lattice is that in the iteration process the interaction
becomes almost harmonic \cite{KMM}, the anharmonic corrections
becoming smaller and smaller as the number of iterations, which is
set by the gasket order $N$, increases. Moreover, for $f\neq 0$
the resulting effective interaction is no longer an even function
of $\theta_{ij}$ \cite{KMM}.

\begin{figure}[tb]
\vbox{ \centerline{
\psfig{file=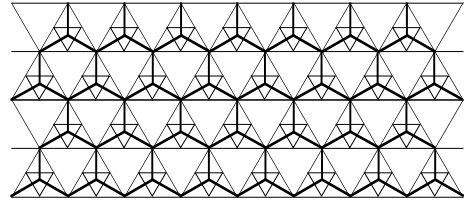,width=0.7\columnwidth,clip=}} \vspace{3mm}
\caption{Successive applications of the triangle-star
transformation allow to replace each gasket by an elementary
"star", thereby turning a lattice of periodically repeated gaskets
into a honeycomb lattice.} \label{fig10} }
\end{figure}

After establishing that the ground
states in the euclidian regime ($f<f_{cN}$) are formed by vortices
occupying the sites of a triangular lattice with concentration
$f_{H}=f/f_{cN}<1$, it is natural to expect that the states which
are particularly robust against thermal fluctuations correspond to
values of $f_{H}$ allowing the formation of an undistorted
triangular vortex lattice (analogous to the Abrikosov lattice in
bulk superconductors) commensurate with the underlying lattice
provided by the array. The energy required to create a defect in
these highly symmetric states is larger than for frustrations
requiring the vortex lattice to be distorted. Therefore, these
Abrikosov-like states will be less vulnerable to thermal fluctuations.
For $1/2<f_{H}<1$ almost equally stable states can be constructed when
the vacancies in the densely packed triangular vortex lattice corresponding
to $f_{H}=1$ also form an undistorted triangular lattice. This leads to
the symmetry $f_{H}\leftrightarrow 1-f_{H}$, although it is not rigorous.

It is readily seen that triangular vortex lattices commensurate
with the underlying lattice can be constructed for
$f_H=1/(m^2+mn+n^2)$, where $m$ and $n$ are integers ($m\geq 1$,
$0\leq n \leq m$). In particular, $m=n=1$ gives $f_H=1/3$, $m=2$
and $n=0$ give $f_{H}=1/4$, $m=2$ and $n=1$ give $f_H=1/7$, {\em
etc.} The states for $f_{H}=1/3$ and $f_{H}=1/4$ are shown,
respectively, in Figs. 11(a) and 11(b). Within this family, the
most "dense" state, \textit{i.e.} the state for $f_{H}=1/3$, can
be expected to be, on account of its stronger vortex-vortex
interaction, the most stable one, followed, in order of decreasing
stability, by those for $f_{H}=1/4$ and $f_{H}=1/7$. This is
precisely what we observe in the magnetoimpedance data of Fig. 9
for the array of second-order gaskets where, focusing on the
interval $|f_{H}|<1/2$), we find that the most prominent
structures, signaling a very stable commensurate state, appear at
$|f_{H}|=1/3$. Weaker structures corresponding to the next-stable
triangular vortex lattice are also well resolved at $|f_{H}|=1/4$,
whereas structures at $|f_{H}|=1/7$ are barely visible and present
only in the inverse magnetoinductance data. It is worth mentioning
that the observation of the particularly prominent vortex state at
$|f_{H}|=1/3$ is entirely consistent with theoretical predictions
\cite{K,SS} as well as with results of Monte Carlo simulations
\cite{SS} for the frustrated $XY$ model on the honeycomb lattice.

\begin{figure}[tb]
\vbox{ \centerline{
\psfig{file=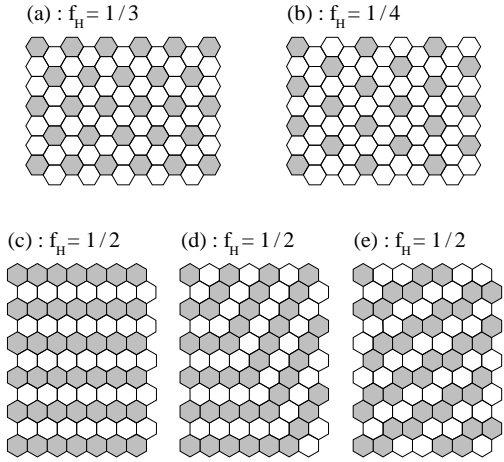,width=0.8\columnwidth,clip=}} \vspace{1mm}
\caption{(a) and (b) show ground state Abrikosov-like triangular
vortex configurations in a honeycomb lattice at, respectively,
$f_{H}=1/3$ and $f_{H}=1/4$. (c), (d), and (e) are examples of
ground-state vortex configurations with the same energy in a
honeycomb lattice at $f_{H}=1/2$. (c): regular 1D superlattice
structure ("striped" phase); (d): a zero-energy domain wall
separating two "striped" states of the type shown in (c); (e): the
state obtained from (c) by introducing the largest possible number
of zero-energy domain walls.} \label{fig11} }
\end{figure}

The magnetoimpedance data of Fig. 9 show weak structures also at
$|f_{H}|=1/2$. The ground states of the $XY$-model on a honeycomb lattice
at this particular frustration and the corresponding vortex configurations
were studied in Ref. \onlinecite{K} and shown to possess a so-called
accidental (\textit{i.e.}, not related to the symmetry) degeneracy, which
can be discussed in terms of zero-energy domain walls. Fig. 11(c)
shows an example of a ground state for $f_{H}=1/2$ characterized
by a regular 1D superlattice structure. In this state all the
variables $\{\theta_{ij}\}$ take the values $\theta_{ij}=0,\pm\pi/4$.
Redistributing the same set of variables
in a different way among the bonds of the lattice allows to
transform the state of Fig. 11(c) into another one with the same
energy, shown in Fig. 11(d), where a domain wall separates two
"striped" states of the type shown in Fig. 11(c). An infinite
family of states with the same energy can be constructed by
creating sequences of such zero-energy domain walls parallel to
each other. Fig. 11(e) shows another example of a periodic ground
state, which can be obtained by introducing the largest possible
number of domain walls into the state shown in Fig. 11(c).

It is known \cite{KVB} that in systems allowing the formation of
zero-energy domain walls the stabilization of long-range order at
nonzero temperatures is achieved if the accidental degeneracy is
removed by thermal fluctuations.  This mechanism (the
so-called "order by disorder" \cite{obd}) is relatively weak and,
therefore, the ordering in frustrated $XY$-models with accidental
degeneracy should be less stable (and destroyed at lower
temperatures) than in other $XY$ models. This effect should be
even more pronounced for the fully frustrated (\textit{i.e.} for
$f_H=1/2$) model on a honeycomb lattice, which, in contrast to
other 2D $XY$-models with accidental degeneracy
\cite{KVB,KK,H,FCKF}, does not display any difference in the
spin-wave free energy, calculated in the harmonic approximation,
of different periodic ground states \cite{K}. The Monte-Carlo
simulations of Shih and Stroud \cite{SS} have indeed demonstrated
that the phase transition of the frustrated $XY$-model on a
honeycomb lattice at $f_H=1/2$ takes place at a much lower
temperature than at $f_H=1/3$ or $f_H=1/4$, where the
accidental degeneracy is absent.

Returning to the system of
periodically repeated SGs at $f=f_{cN}/2$ ($f_H=1/2$), it is also
possible to start the search for its ground state by finding the
structure of the state which minimizes the energy (\ref{E}) for
the periodic vortex configuration shown in Fig. 11(c). Obviously,
this state is periodic, and its unit cell comprises two gaskets. By
redistributing the same set of $\{\theta_{ij}\}$ among the bonds of the
lattice in a different way, one can construct the state with the
same energy corresponding to the vortex configuration shown in Fig. 11(d).
However, in contrast to the fully frustrated $XY$-model on a
honeycomb lattice, the state obtained in this way will not be an
extremum of the Hamiltonian and, therefore, a slight readjustment of
$\{\theta_{ij}\}$ can further decrease its energy. This
means that in a system of periodically repeated SGs the domain
wall shown in Fig. 11(d) has a negative energy. Accordingly, the
state with the lowest energy corresponds to the periodic vortex
configuration shown in Fig. 11(e), which is characterized by the
highest possible density of such domain walls. The unit cell of
this state comprises four gaskets. In the fully frustrated
$XY$-model on a honeycomb lattice a ground state with the same
structure is selected if one takes into account the interaction,
of arbitrary sign, with the second-nearest neighbors \cite{KU}.

Thus, the phase transition taking place, with decreasing temperature, at
$f_H=1/2$ should be related to the appearance of long-range order corresponding
to the vortex configuration shown in Fig. 11(e). The selection of
this state relies on a weak mechanism, which loses its efficiency with
increasing $N$, since the effective interaction becomes almost harmonic under
decimation. One can therefore expect the ordered phase at $f_H=1/2$ to
be again rather vulnerable to thermal fluctuations. This explains why
the structures at $|f_{H}|=1/2$ in Fig. 9 are much weaker
\cite{comm} than those at $|f_{H}|=1/3$. Their strength is at most
comparable to that of the structures corresponding to the
triangular vortex lattice at $f_{H}=1/4$, whose weaker
vortex-vortex interaction makes the ordering less robust than at
$f_{H}=1/3$. Numerical simulations \cite{SS} show that in the
conventional $XY$ model {on a} honeycomb lattice the ordered
phases appear, with decreasing temperature, in the same order:
{first} at $f_H=1/3$, {then} at $f_{H}=1/4$, and only {further
down} at $f_H=1/2$.

\section {Conclusion}

Our magnetoinductance measurements on Josephson junction arrays of
periodically repeated Sierpinski gaskets have clearly demonstrated
the existence of two regimes. In agreement with theoretical
analysis, in one of them (the fractal regime) the peaks observed
in the inverse sheet magnetoinductance, reflecting those states
which are the most stable against thermal fluctuations, are
equally spaced. Neighboring states in this sequence differ from
each other by the penetration of an additional vortex into each
unit cell of the array. In the other, euclidian, regime the
sequence of the observed stable states corresponds to periodic
lattices of vortices occupying the largest triangular cells of the
array.

The agreement with theory is
achieved not only for the positions of the different peaks, but
also for their relative strengths. In the euclidian regime the
structure of the ordered states is analogous to that on a
honeycomb lattice with reduced frustration and the relative
stability of the different states can be understood in terms of
vortex lattice disordering. However, the amplitudes of the
peaks observed in the fractal regime can be quantitatively
explained only if the redistribution of the magnetic field in the
array due to the asymmetric shape of the superconducting islands
is taken into account. In this connection, we would like to
mention the recent work by Park and Huse \cite{PH}, who compared
the energies of different states in a wire network with a {\em
kagom\'{e}} lattice geometry at full frustration. These authors
came to the conclusion that the effects related to the finite
width of the wires can be compensated by {\em bending} the wires.
This is equivalent to our conjecture that the influence of the
asymmetry associated with the screening currents can be reduced to
a redistribution of the magnetic field with respect to an ideal
system.

The results of our magnitoinductance measurements show
that this phenomenon is more pronounced at high frustration
levels, an observation consistent with the analysis presented in
Sec. III B. Indeed, in our model, the triangular cells exhibiting
the largest relative areal changes are the smallest (elementary)
ones, which are precisely those providing the dominant
contribution to $L^{-1}(f)$ at high values of $f$.  Data taken at
very small frustrations, like those shown in Fig. 9, are
practically unaffected by the nonuniform frustration resulting
from the asymmetry of the screening currents.

Early studies \cite{GG3} of a large-order ($N=10$) single gasket of
superconducting aluminum wires revealed that the period associated with
adjacent minima of the superconducting-to-normal phase boundary $T_{c}(f)$
was larger than that extracted from adjacent maxima. We have found
a similar perturbation of the periodic field dependence in our
calculations of $L^{-1}(f)$ based on the model for the magnetic
field redistribution proposed in Sec. III B. The same mechanism
may be responsible for the anomalous feature observed in Ref.
\onlinecite{GG3}. It should be noticed, however, that the
experiments of Ref. \onlinecite{GG3} were performed at
temperatures very close to $T_{c}(0)$, where the estimated
magnetic penetration depth of the Al wires turns out to be
comparable to their width. Accordingly, the effective gasket
distortion resulting from screening effects is expected to be
weaker than in our arrays. A quantitative verification is therefore needed
before drawing a conclusion as for the ability of our model to explain
the anomaly observed in Ref. \onlinecite{GG3}.

The influence of the incommensurability of different lattice cells on the
magnetoresistance of Josephson junction arrays was discussed
by Kosterlitz and Granato \cite{KG} in relation to experiments
performed on periodic arrays with a complex unit cell \cite{VW}.
However, quantitative agreement between experiment and theory in
treating incommensurability effects in Josephson junction arrays
has been demonstrated only by the present work \cite{comm2}. In
our system the phenomenon of incommensurability is present in an
hidden form and its manifestations appear as a consequence of the
asymmetric shape of the superconducting islands.

\acknowledgments

We are grateful to H. Beck for several interesting discussions. We also
thank R. Th\'{e}ron for his assistance in preparing the figures. P.M.
would like to thank {\O}. Fischer and J.M. Triscone for the hospitality
extended to him during an eight-month visit at the DPMC of the University
of Geneva, where part of this article was written. This
research was supported by the Swiss National Science
Foundation, the Swiss Federal Office for Education and Science
within the framework of the TMR-project "Superconducting
Nanocircuits" of the European Union, and the "Vortex Scientific
Program" of the European Science Foundation. S.E.K.  additionally
acknowledges the support of the Program "Quantum Macrophysics" of
the Russian Academy of Sciences and of the Program "Scientific
Schools of the Russian Federation" (grant No. 00-15-96747).


\begin{references}
\bibitem{comm2}Concerning periodic wire networks with incommensurate cells,
               the agreement between experimental investigations
               and theoretical calculations of the SN phase
               boundary $T_c(B)$ was established in Refs.
               \onlinecite{BSN}.
\end{references}
\end{document}